\documentclass[10pt, aas2pp4]{aastex}

\shorttitle{Bright Optical Transients}
\shortauthors{Shamir \& Nemiroff}

\begin{document}

\title{Frequency Limits on Naked-Eye Optical Transients Lasting from Minutes to Years}

\author{Lior Shamir}
\affil{Image Informatics and Computational Biology Unit, Laboratory of Genetics, NIA, NIH, 333 Cassell Dr., Suite 3000, Baltimore, MD 21224}
\email{shamirl@mail.nih.gov}

\and

\author{Robert J. Nemiroff}
\affil{Department of Physics, Michigan Technological University, Houghton, MI 49931}
\email{nemiroff@mtu.edu}

\begin{abstract}
How often do bright optical transients occur on the sky but go unreported?  To constrain the bright end of the astronomical transient function, a systematic search for transients that become bright enough to be noticed by the unaided eye was conducted using the all-sky monitors of the Night Sky Live network.  Two fisheye continuous cameras (CONCAMs) operating over three years created a data base that was searched for transients that appeared in time-contiguous CCD frames.  Although a single candidate transient was found (Nemiroff and Shamir 2006), the lack of more transients is used here to deduce upper limits to the general frequency of bright transients.   To be detected, a transient must have increased by over three visual magnitudes to become brighter than visual magnitude 5.5 on the time scale of minutes to years.   It is concluded that, on the average, fewer than 0.0040 ($t_{dur} / 60$ seconds) transients with duration $t_{dur}$ between minutes and hours, occur anywhere on the sky at any one time.  For transients on the order of months to years, fewer than 160 ($t_{dur} / 1$ year) occur, while for transients on the order of years to millennia, fewer than 50 ($t_{dur}/1$ year)$^2$ occur.
\end{abstract}

\section{Introduction}

Transients that become bright enough to be noticed with the unaided eye have been recorded on the night sky far back into human history.  Ancient records include, for example, bright supernovae such as the SN 1054 which reached an estimated visual magnitude of $-4$ and created the Crab Nebula \citep{Duy42, May42}.  Historically, naked-eye transients have helped inspire peoples to stop wars and topple governments \citep{Gol06}.  Recent examples of transients reaching naked-eye brightness include the bright Comet McNaught, which reached a visual magnitude near $-6$ \citep{McN07} in 2007 January, and Comet Holmes, which sustained a brightness near visual magnitude three for months near the end of 2007 \citep{Mor08, ElH09}.  Most recently a gamma-ray burst GRB 080319B reached naked eye brightness for on the order of one minute \citep{Cwi08, Blo08}.

Several different types of known, extra-terrestrial, astronomical phenomena can create transients that become bright enough for naked-eye detection.  These include novae \citep{Sha97}, supernovae \citep{Kun87}, optical counterparts to gamma-ray bursts \citep{Nem04, Cas01, Ofe03, Blo08}, microlensing \citep{Col95, Nem98}, stellar superflares \citep{Sch00}, and comets \citep{Hug90}.   

Alternatively, there are historical claims of optical transients that are not clearly matched to known astronomical phenomena, such as the now discredited Perseus Flasher \citep{Kat86, Sch87}, which reportedly kept flashing near visual magnitude $-1$ at durations of less than one second.  Conceivable, empirically discovered transients might not match up with any progenitor mechanism yet known, and therefore be particularly interesting.

In the past few years, several robotic telescopes have been deployed to search for optical transients of various magnitudes \citep{Cer03, Ake03, Ves04, Ste04, Lip04, Sat04}.  Primary research goals of such telescopes include the discovery of supernovae and the optical counterparts to gamma-ray bursts.  Bright transients detected originally in many wavelength bands have been reported in Circulars of the International Astronomical Union coordinated by B. Marsden, the Gamma ray bursts Coordinate Network (GCN) coordinated by S. Barthelmy, and in the email based Astronomer's Telegram \citep{Rut98}.   

This paper focuses on a certain type of optical transient -- one that unexpectedly brightens by more than an order of magnitude during an observing campaign to become visible to the unaided eye.  This does not include, for example, Mira variables, as they are well known and predictable.  This also does not include transients created by planets eclipsing stars, as those would typically involve only a small change in the overall source's brightness.  Optical transients that reach naked-eye brightness are quite rare.  Detecting these transients could prove quite valuable, however, as they are discoverable and trackable with even modest cameras.   

One approach of searching for bright optical transients is by deploying continuously operating, panoramic, all-sky cameras.  The faintness limits on many all-sky cameras can be made similar to the unaided human eye \citep{NeS03, NemToS03, Nem04, Pic06}.  Continuous operation increases the chance that optical transients appearing on clear nights are recorded \citep{Nem99}.  The most comprehensive fisheye sky monitoring system to date was Night Sky Live network of all-sky cameras \citep{NeS03}.  At its peak in 2005, the first Night Sky Live network deployed eleven fisheye cameras simultaneously recording most of the night sky brighter than visual magnitude five, most of the time.  

In this paper, data taken from three years of night sky coverage by two cameras in the {\it Night Sky Live} network of all-sky cameras \citep{Nem99} is used to derive limits on the frequency of naked-eye transients.  Section~\ref{theory} reviews how the results of transient searches can be converted to occurrence limits.  Section~\ref{observations} describes the observations, hardware, and software used to search for naked-eye transients, and present the results of the search.  Section~\ref{limits} converts the search results to fundamental limits on the frequency of bright astronomical transients.  Section~\ref{conclusions} summarizes, gives discussion, and draws conclusions.

\section{Theory: Connecting Transient Searches to Transient Limits}
\label{theory}

To connect results of a transient search to actual limits on the event rates of transients is straightforward but sufficiently nuanced that a brief analysis is presented here.  In sum, a hypothesis involving the abundance of transients is made and statistically compared with the observational result.  In this simple canonical model, a transient instantly brightens by more than a given magnitude amount from the beginning of an observing campaign.  The transient then does not dip below this peak brightness for its duration $t_{dur}$.  A simple canonical observing campaign of duration $t_c$ is also assumed where consecutive exposures all have exposure duration $t_e$.  For a discussion on detecting transients that includes down time, optimal frame rates and statistical independence, see \citet{NemToS03}.  

The standard hypothesis that will be tested is that $N_i$ instantaneously detectable transients of duration $t_{dur}$ occur randomly on the sky at any one time.  Therefore, most simply, a hypothetical device that images the entire sky would enable the detection, on the average, of $N_i$ transients on every all-sky image.  A more realistic device would have an angular efficiency factor of $f_a$ and a temporal efficiency factor of $f_t$.  Here $f_a$ is the fraction of sky that is being monitored, while $f_t$ is the fraction of time during the campaign that a transient could actually be detected.  Variable names and nomenclature will follow \citet{NemToS03}.

As discussed in some detail below, for given search parameters and transient durations $t_{dur}$, the statistically independent number of trials $N_{trial}$ might differ from number of exposures taken during the observing campaign $N_c$.  The relationship between $N_{trial}$ and $N_i$ may involve $f_t$, $t_{dur}$, $t_e$, or even $t_c$.  Once $N_{trial}$ is determined, however, then the grand number of transients $N_g$ detected during the observing campaign would be 
 \begin{equation} \label{NgNi}
 N_g = f_a N_{trial} N_i .
 \end{equation}

Note that it is not only the detection of light from transients themselves that is important here.  Rather, what is searched for is the {\it change} in the brightness of a source.  Otherwise, all detected stars would be classified as transients.

\subsection{Single frame transients}

When each exposure duration $t_e$ in the transient search observing campaign is significantly greater than a given canonical transient duration $t_{dur}$, only ``single frame transients" of that duration are usually expected.  Single frame transients should generally be considered to be unreliable indicators of extra-terrestrial astronomical phenomena, as any number of local effects could mimic them, including cosmic-ray hits, satellite glints, and meteors.  Cosmic ray hits could be ruled out, however, based on image shape \citep{Sha05e} or were several cameras to record a transient independently.  Satellite glints \citep{Nem05} and meteors could be ruled out with simultaneous detections of sufficient parallax so as to create a minimum limiting distance of extra-terrestrial origin.  Alternatively, a group of single frame transients might be believable as a statistical ensemble.

Mathematically, for transients with $t_e >> t_{dur}$, transients visible on one frame would not be expected to be visible on any other frame.  Therefore, for the purposes of statistics, a search for transients on one frame should be considered statistically independent of a search on any other frame.  Each frame could be considered an independent statistical ``trial" in a search for transients.  Therefore, for single frame transients, effectively $f_t = 1$ and $N_{trial} = N_c$.  Therefore, Eq. (\ref{NgNi}) for the grand total number of transients becomes 
 \begin{equation} \label{Ng1}
 N_g = f_a N_c  N_i .
 \end{equation}

Note that when $t_e >> t_{dur}$, a transient is not ``on" during the entire frame or frames when it is detected.  Therefore, these fast transients would have to be brighter during their short reign to appear as bright as a seemingly similar quiescent source in the same frame \cite{Nem98b}.  Given a limiting detectable brightness of a quiescent source of $l_{dim}$ per frame, and assuming the transient occurs during a single frame, the limiting brightness of a transient source would need to be $l_{transient} > l_{dim} (t_e / t_{dur})$ to be detected above the quiescent limit.

\subsection{Multiple frame transients}

When a single transient is detectable on several consecutive frames, each frame should {\it not} be counted as a statistically independent trial.   This is because a single long duration transient is much more likely to occur on successive frames than two independent transients.  For time consecutive frames, the effective number of statistically independent trials is $N_{trial} \sim N_c (t_e / t_{dur})$.   More generally, given the hypothesis that on average $N_i$ transients exist instantaneously somewhere on the sky at any one time, the grand total of detected transients expected during an observational campaign is expected to be, given perfect time coverage, $N_g = f_a N_{trial} N_i \sim N_c f_a N_i t_e / t_{dur}$.  

In many practical searches, however, there are time lapses in time coverage due to clouds, daylight, or exposures missing for other reasons.   In these cases the effective number of statistical trials is reduced by a ``temporal efficiency" factor quantized as $f_t$, the fraction of time during a campaign that, were a transient to start, it would be detected during the campaign as a transient. 

It is important to note that different $f_t$ factors may operate for different transient durations $t_{dur}$.  Consider, for example, a one week observing campaign.  Further consider that due to clouds, every other night is completely missed.  For transients with $t_{dur} \sim 1$ day, the efficiency of exposures during the week would be $f_t = 3 / 7$.  For transients durations on the time scale $t_{dur} \sim 1$ week, however, a different $f_t$ fraction determines the effective number of trials.   This is because a one week transient that occurred any time during the week long observing campaign would have been detected with near unit efficiency.  Therefore, for week long transients in this example, $f_t \sim 1$.  More generally, $N_{trial} \sim f_t N_c t_e / t_{dur}$ .

For believability, an observer might demand that a ``verified" transient occur on (at least) $N_{ver}$ frames, quite possibly time-consecutive frames.  This would primarily affect the verified detection of transients near the exposure time of individual frames.   As in the above case, this would reduce the effective number of statistically independent trials by a factor of $N_{ver}$.  Therefore, the effective number of trials would be
 \begin{equation} \label{NtM}
 N_{trial} = { f_t \ t_e  \ N_c \over
                               N_{ver} \ t_{dur} } .
 \end{equation}
Also, the grand total number of verified transients of duration $t_{dur}$ expected during the observing campaign would then be derived from Eq. (\ref{NgNi}) to be 
\begin{equation} \label{NgNiM}
 N_g = \left( { f_a \ f_t \ t_e  \ N_c \over
                                         N_{ver} \ t_{dur} } \right)  N_i  .
 \end{equation}

The situation is slightly more complicated were frames co-added.  Assume that $N_{add}$ frames are co-added so that a transient is {\it only} visible on the co-added frames -- it is too dim to be significantly detected otherwise.  In this case, the effective number of independent trials would again be reduced, and the resulting expressions can be found by substituting $N_{ver}$ with $N_{ver} N_{add}$ in the above expressions.

\subsection{Transients lasting longer than the observing campaign}

Limits on $N_i$ can even be determined for transient durations $t_{dur}$ greater than the entire duration of the observing campaign $t_c$.  The important feature for detection here is that the transient's rise in brightness occurs {\it during} the observing campaign.  It will be assumed here that a transient that declines in brightness will not be found because it will not be searched for.  

Consider as an example a one week observing campaign's effort to detect a transient with a one year duration, such that $N_i = 1$, meaning that such a transient is likely to occur somewhere on the sky at any particular time. On average one measurable event will occur during that year -- the rise of the transient.  The chance that the rise occurs during any one week during that year is about 1 in 52, so that, effectively, for transients of $t_{dur} \sim 1$ year, $f_t \sim 1 / 52$.  

More generally, the effective number of trials $N_{trial}$ can be convolved with an $f_t$ that dips to less than one such that 
 \begin{equation} \label{NtC}
N_{trial} = f_t = t_c / t_{dur}.  
 \end{equation}
Therefore Eq. (\ref{NgNi}) becomes 
 \begin{equation} \label{NgC}
 N_g =  f_t f_a N_i = { \left( f_a t_c  \over  
              t_{dur} \right)} N_i .  
 \end{equation}

\subsection{Frequency Limits from Null Detections}

What is the greatest number of instantaneous transients $N_i$ that is consistent with no transients being found ($N_{act} = 0$) during the observational campaign?  To find $N_i$, one might start with the intermediate question: How different is this $N_g$ from zero?  Now the number of $\sigma$ that $N_g$ differs from zero is $\sigma = \sqrt{N_g}$.  Therefore, the difference between $N_g$ and zero in terms of $\sigma$ is 
 \begin{equation}
  s = { (N_g - 0) \over \sqrt{N_g} } = \sqrt{N_g}  .
 \end{equation}
Choosing $s$ as quantifying the upper limit on how high $N_g$ can become gives
 \begin{equation}
  N_g < s^2  .
 \end{equation}
Given now the relation Eq. (\ref{NgNi}) this becomes
 \begin{equation} \label{Ninull}
 N_i < { s^2 \over
            f_a N_{trial} } .
\end{equation}

Now $N_{trial}$ can be expanded for each of the three cases discussed above.  For the single frame transient, $N_{trial} = N_c$ so that 
 \begin{equation} \label{NiS0}
 N_i < { s^2 \over f_a N_c}  .
 \end{equation}
For the multiple frame transient, one substitutes Eq. (\ref{NtM}) into Eq. (\ref{Ninull}) to get
\begin{equation} \label{NiM0}
 N_i <   { N_{ver} \ t_{dur} s^2 \over
                                     N_c \ f_a  \ f_t \ t_e }    .
 \end{equation}
For transients longer than entire observing campaign, one substitutes Eq. (\ref{NtC}) into Eq. (\ref{Ninull}) to get
\begin{equation} \label{NiC0}
  N_i < { t_c \ s^2  \over  f_a \ t_{dur} } .
  \end{equation}

\subsection{Frequency Limits Given Actual Detections}

What is the greatest number of instantaneous transients $N_i$ that is consistent with the results of a campaign that observed $N_{act}$ actual transients?  To find out, the logic of the above subsection is followed focusing on the question: How different is this $N_g$ from $N_{act}$?  As above, the standard deviation of the predicted $N_g$ is $\sigma = \sqrt{N_g}$.  Therefore, the difference between $N_g$ and zero in terms of $\sigma$ is 
 \begin{equation}
  s = { (N_g - N_{act}) \over \sqrt{N_g} }   .
 \end{equation}
Choosing $s$ as quantifying the upper limit on how high $N_g$ can become gives
 \begin{equation}
  s^2 > { (N_g  - N_{act})^2 \over N_g }.
 \end{equation}
This equation is quadratic in $N_g$ and has a physical solution for $N_{act} > 0$ such that 
\begin{equation} \label{NgNa4}
 N_g < (N_{act} + s^2/2) +  \sqrt{N_{act} s^2 + s^4/4}  .
 \end{equation}
Substituting in Eq. (\ref{NgNi}) into the above equation yields
 \begin{equation}  \label{NiNtrial}
 N_i < { (N_{act} + s^2/2) +  \sqrt{N_{act} s^2 + s^4/4} \over
              f_a \ N_{trial} } .
 \end{equation}

Single frame transients are again first considered. Plugging in $N_{trial} = N_c$ for single frame transients into Eq. (\ref{NiNtrial}) yields
\begin{equation} \label{NiS1}
 N_i < { (N_{act} + s^2/2) +  \sqrt{N_{act} s^2 + s^4/4} \over
              f_a \ N_c } .
\end{equation}
For multiple frame transients, using the $N_{trial}$ of Eq. (\ref{NtM}) yields
\begin{equation} \label{NiM1}
N_i <   \left( { N_{ver} \ t_{dur} \over f_t \ t_e  \ N_c } \right)
            [  (N_{act} + s^2/2) +  \sqrt{N_{act} s^2 + s^4/4} ].
 \end{equation}
For transients with durations greater than the observing campaign, using the $N_{trial}$ of Eq. (\ref{NtC}) yields
 \begin{equation} \label{NiC1}
 N_i <  \left( {t_{dur} \over t_c}  \right)
             [ (N_{act} + s^2/2) +  \sqrt{N_{act} s^2 + s^4/4} ] .
 \end{equation}

\section{Observations: A Three Year Search for Bright Astronomical Transients}
\label{observations}

\subsection{Hardware}

Two CONtinuous CAMeras (CONCAMs) from the Night Sky Live network were used for this project.  These CONCAMs were located in Cerro Pachon, Chile, and the Canary Islands, off the west coast of northern Africa.  Although other CONCAMs were running, transient detection software was only applied to data taken by these two cameras.  Reasons that other cameras were not used include bandwidth, hardware, and remote computing power limitations.   Each CONCAM includes a CCD camera, a fisheye lens and an industrial PC running Linux Red-Hat. The CCD cameras used were SBIG ST-1001E. The lenses used were the SIGMA F4-EX.  However, clouds, occlusions, and reduced sensitivity very near the horizon limited the search to approximately $\pi$ steradians per clear frame.  

The CONCAM hardware would typically run continuously from nautical sunset to nautical sunrise.  The resulting images are 1024 $\times$ 1024 pixels, with approximately $2^{16}$ gray-scale levels and recorded in FITS format.  Each CONCAM recorded a 180-second exposure every 236 seconds.   The reason for the 180 exposure duration related to the time before stars began to trail significantly on the image.  The reason for the 236-second time between the beginnings of exposures is to allow for data read-out and to allow each CONCAM to begin exposures separated in time by exactly one sidereal day.  The reason for separating exposures by exactly one sidereal day is to allow stars to return to known positions in the sky and on the CCD, allowing fewer and more manageable systematic effects.  

The CONCAM all-sky cameras were passive and did not track the sky.  The wide-angle lenses allow recording full 2$\pi$ steradians in one frame, and quiescent stars as dim as visual magnitude 6.8 are visible near the image center \citep{Sha05a}.   

\begin{table}
\begin{center}
\begin{tabular}{lc}
\hline  Color & Visual Magnitude \\
\hline 
A & 5.1 \\
F & 5.2 \\
G & 5.3 \\
K & 5.6 \\
M & 5.8 \\
\hline
\end{tabular}
\caption{CONCAM stellar visual magnitude of a 40$\sigma$ PSF}
\label{lim_mag}
\end{center}
\end{table}

\subsection{Software}

Once taken, the images were transmitted to a central computer at Michigan Technological University, where they were made publicly available over a web server accessible at http://nightskylive.net/ .  Data for this project were then analyzed for new astronomical transients.  

The detection of optical transients in all-sky images requires several logical steps.  Before even the first transient was sought, it was useful to build a database of canonical images that are known {\it not} to have a transient that can then be compared to images that might contain a transient.  
For this project, this was done by co-adding several images taken on a clear night at the same sidereal time. A detailed description of this mechanism is described in \citep{Sha05b}.

Each source that is detected above 40$\sigma$ on a patrol frame is compared to the same source position on the comparison frame.  A threshold of 40$\sigma$ was found to give a resultant rate of candidate single-frame transients that could be checked by humans. If that source is not detected on the comparison frame, meaning specifically that no source above 2$\sigma$ is detected there, then it has passed the first cut and is a candidate transient.  

Note that the high 40$\sigma$ change insures a minimum amount of variability in actually detected transients.  Surely every source is variable at some low level, but the transient sources searched for here must be variable above some minimum level to be detectable here.  What is that level?  Given a background level of 1000 counts, a typical level, 40$\sigma$ over background corresponds to about 1200 additional counts, or about 2200 raw counts.   Assume a quiescent source was as bright as 2$\sigma$ over background previously and had gone undetected.   The source would then have undergone a change from 64 to 1200 counts.  This corresponds to a factor of about 18.75, or a variability of about three magnitudes at minimum.

After the comparison frame cut, any surviving candidate transient is then searched for in an online catalog.  The catalog used for comparison was the Hipparcos catalog \citep{Per97}.  In this way, several candidate transients were later identified as variable stars.  

Even after the catalog comparison cut, most candidate transients turned out to be single frame events that were not astronomical in origin.  There are many routes to creating single-frame ``background" transients that may initially appear to be astronomical in nature.  Hot stuck pixels, cosmic rays, Moon-ghosts, satellite glints all provided a background for single frame transients, while planets provided a background for multiple frame transients.  

One important step in rejecting background single-frame events was the rejection of pixels dominated by cosmic-ray generated counts \citep{Sha92,Fix00}, typically made use of the non-point source nature of cosmic ray splashes \citep{Sha05c}.  Bright planets are also rejected using a star recognition algorithm designed to find astronomical objects in wide angle frames \citep{Sha05d}. 

Due to the relatively high density of artificial objects in orbit around the earth, one can expect that many single-frame candidate transients are actually sun glints from these artificial objects \citep{Sch87a,Var92}. In fact, some flashes that were suspected to be true astronomical transients \citep{Hal87} appeared later to be nothing but foreground near-earth flashes \citep{Sch87}. Another source of background flashes is bright meteors and fireballs \citep{Sha05c}. When the trajectory of a meteor is oriented toward the camera, the meteor might have a point spread function (PSF) that appears similar to that of a star \citep{Bro02}.

Even given these rejections, the non-astronomical background remained so high that little trust was placed on the hypothesis that any transient that appeared on a single frame was astronomical in origin.  Therefore, subsequent cuts by the software demanded that candidate transients be significantly brighter than frame background and were found in more than one consecutive CONCAM frame.  Frames taken at the same time from other CONCAMs were compared as well, but this step was only used to confirm candidate transients.  

Although the CONCAM exposure duration was chosen to be just long enough so that most stars don't trail significantly, in fact most star centroids do trail by a few pixels each exposure.  This seemingly disadvantage was used here to find transients that rotated with the sky. If the PSF of the flash seems to trail to the same direction of nearby stars, this is an indication that the transient had little angular motion relative to these stars, and so the trail might be caused by the rotation of the Earth.  Flashes with non-trailing PSFs were interpreted as either very short flashes or flashes from transients that rotated with the Earth, such as geosynchronous satellites \citep{Sha05e}.

Once a flash is recorded it is compared to previous images to check if the flash is persistent and rotates with the sky. If a flash appears to be rotating with the sky for at least two images, the system alerts on that flash as an optical transient candidate \citep{Sha06}.

\subsection{Angular Efficiency}

To estimate the angular fraction of the sky $f_a$ visible during each frame, it is noted that each {\it CONCAM} all-sky camera sees the entire sky over the horizon, which amounts to $2 \pi$ steradian.  However, the area nearest the horizon has low visibility, and the automated source search only operated at 20 degrees over the horizon.  This means that about 34 percent of the sky is ignored, yielding an effective search area, per CONCAM, was about 4.1 steradians.  Together, the two CONCAMs had a higher search area, but for sake of conservative limits the total search area will be considered to be 4.1 steradians.  Dividing by the total angular area of the sky, $4 \pi$ steradians, it is found that the total fraction of sky monitored per exposure is about $f_a \sim 4.1/(4 \pi) \sim 0.33$.

\subsection{Temporal Efficiency}

The transient detection mechanism started operating in October 2003 in La Palma, Canary Islands, and another station started operating in Cerro Pachon, Chile in September 2004.  

The present search was not uniformly sensitive to transients of all durations.    Here the sensitivity of the search to transients of all durations $t_{dur}$ will be assessed.   The search sensitivity to transients of the shortest durations will be evaluated first: those lasting less than 12 minutes.  These might be seen on individual campaign frames but could {\it not} exist on three consecutive and independent frames.  Transients this short would therefore be considered unverified and hence not reported as a discovery, so that $f_t \sim 0$.  

Next, the temporal efficiency $f_t$ is evaluated for a transient with $t_{dur}$ of 12 minutes.  A 12 minute transient would likely not be missed were it to occur during a clear, moon-free time when a CONCAM was operating.  On the average, CONCAMs run 8 hours during any 24 hour period, which is $0.33$ of a day.  Also, on the average, CONCAMs are able to record transients only about one night in four, due to weather, bright moon time when the software is not running, and equipment problems \citep{Per05}. Therefore, were a 12 minute transient to occur in the field of a CONAM during this three year campaign, the chance of it actually being recorded is about $f_t \sim 0.33/4 \sim 0.0825$.  

Next, a transient with a duration of one hour is considered.  This situation is very similar to the 12 minute transient, and the estimated efficiency factor would be $f_t \sim 0.33/4 \sim 0.0825$.   The same logic holds true for transients of duration 8 hours.  

Next, let's consider a transient with a duration of one week.  Here the situation is a little different.  The fact that CONCAMs only operate for on average 8 hours a day becomes irrelevant.  This is because a CONCAM operating only 8 hours is sure to catch a one-day transient in at least three of its many independent exposures during the night.  Therefore, the $0.33$ of a day factor should be excluded.  The other efficiency factor, that CONCAMs obtain useful data only one night in four, is still important, however.  Therefore, for the one day time scale, $f_t \sim 0.25$.

Next, let's consider a transient with a duration of one month.  Here the situation is again different. Here the fact that CONCAMs operate only one night in four also becomes irrelevant.  This is because CONCAMs certainly run at least one night every month, and during this night there would surely be at least three independent exposures that could detect the transient.  Therefore, for the one month time scale, $f_t \sim 1$.

Next, let's consider a transient with a duration of one year.  Since CONCAMs operated during each year of the three year campaign, surely at least three independent exposures would be carried out that would detect the transient.  Note that even a transient that started the night before the campaign ended would likely be found, so that the sensitivity to transients of one year of this three year campaign is very nearly indeed $f_t = 1$.  Note, however, that a transient that started before the campaign started -- or on the first day of the campaign, might not have been recorded by this campaign because it would have been included in the canonical frames against which transients are compared and hence discovered.  

For transients of durations of $t_{dur} = 3$ years, the logic is the same as for $t_{dur} = 1$ year, so that $f_t \sim 1$.

Next, let's consider a transient of duration of $t_{dur} = 10$ years.  Transients of durations longer than three years last longer than the entire observing campaign.  There is a chance the transient remained constant within the quiescence criteria all during the observing campaign, as so was identified as a star and not a transient.   For this reason, as discussed in an above Section, $f_t$ will be less than unity for transient durations greater than the duration of the observing campaign.  Here $f_t =  t_c / t_{dur} \sim 6 / 10 = 0.3$.   Similarly, for transients of $t_{dur} = 100$ and 1000 years, $f_t$ is equal to 0.03 and 0.003 respectively.   The relation between $f_t$ and $t_{dur}$ is shown in Table 1 and depicted graphically in Figure 1.

\subsection{Number of Exposures}

To estimate the total number of exposures taken during the campaign, $N_c$, it is noted that each CONCAM camera takes on average $\sim$140 images per night.  Since only 25\% of the images are clear and moon-free \citep{Per05}, though, each camera produces $\sim$35 clear images per calendar day \citep{Sha06b}.  For a campaign time of three years, this yields approximately $m \sim$ 38,300.

\subsection{Search Results}
\label{search_results}

Only a single multi-frame transient survived both software detection and vetting by the authors \citep{Nem06a}. This candidate transient lasted about 12 minutes, was seen on three time-consecutive CONCAM frames from the Cerro Pachon CONCAM and the (only) two time-coincident frames from the Canary Island CONCAM.  The event was labeled OT060420 where ``OT" stood for optical transient, and the numbers corresponded to the date the candidate transient occurred.  The transient software described above alerted on OT060420 in near real time and the possibility that the event was a real astronomical transient was considered high enough on initial inspection to release the images and position of the transient immediately to the GCN global network for follow-up observations.  Soon thereafter a second report indicated that the transient was not recorded by a somewhat-inferior third all-sky camera \citep{Sme06} led to the hypothesis that the recorded source was not astronomical \citep{Nem06b}.   The nature of the transient remains controversial, with some astronomers interpreting the event as a new type of astronomical transient \citep{Pac06}. A detailed analysis of this event is described in \citep{Sha06b}.

\section{Fundamental Limits on Bright Optical Transients}
\label{limits}

The below discussed results were obtained by including the above discussed search parameters and efficiencies into Eqs. (\ref{NiM0}, \ref{NiC0}, \ref{NiM1}, \ref{NiC1}).   Limits on the all sky prevalence of optical transients $N_i$ were placed demanding $s = 2 \sigma$ upper limit.  

Table 2 summarizes the limiting maximum number of transients instantaneously visible on the sky at any one time: $N_i$, as a function of transient duration $t_{dur}$.   The first column of Table 2 shows the transient duration $t_{dur}$ to which the current campaign was sensitive.   The second column of Table 2 shows the angular efficiency factor $f_a$, which holds for all transients in the transient search campaign reported here.   The third column of Table 2 shows the temporal efficiency factor $f_t$ for a transient of duration $t_{dur}$ in the observing campaign reported here.  

The fifth column of Table 2 shows $N_i$, the maximum number of transients that can be expected to be on the entire sky at any one time with the duration listed in the first column, given that the single successful candidate reported in \citet{Sha06b} is considered real.  As indicated above, this is a 2$\sigma$ upper limit.  In other words, it is possible to have fewer transients on the sky than this limit, but not more.  The fourth column in Table 2 similarly shows $N_i$, but this time assumes that the observing campaign found no transients, essentially assuming that the candidate transient reported in \citet{Sha06b} was not astronomical in nature.   

\begin{table*}[ht]
\begin{center}
\begin{tabular}{| c | c | c | c | c|}
\hline
Transient & Angular & Temporal & Maximum & Maximum \\
Duration & Efficiency & Efficiency & Transients & Transients \\
(units) & $f_a$ & $f_t$ & $N_i, N_{act}=0$ & $N_i; N_{act}=1$ \\
\hline
12 minutes   &   0.33 & 0.0825  & 0.062 & 0.046  \\
1 hour          &  0.33  &  0.0825 & 0.345 & 0.230   \\
8 hours        &   0.33 &  0.0825  &  2.76   &  1.84   \\
1 week        &   0.33 &   0.25    &  19.1   &  12.8    \\
1 month      &   0.33 &  1.0        &  20.5    &  13.7   \\
1 year         &   0.33 &  1.0        &  250     &  166    \\
3 years       &   0.33 &  1.0        & 749      &   499    \\
10 years      &  0.33 & 0.3       &  8,320   &  5,550   \\
100 years    & 0.33  & 0.03     &  8.32E5 &  5.55E5  \\
1,000 years  &  0.33 & 0.003  &  8.32E7  &  5.55E7  \\
\hline
\end{tabular}
\caption{\label{tab1}A table of transient durations and their corresponding maximum number of transients possible on the sky.}
\end{center}
\end{table*}

The relations between $t_{dur}$ and $N_i$ are shown graphically in Figure 2.  Generalizing the above results, it is concluded that, on the average, less than 0.0040 ($t_{dur} / 60$ seconds) transients with duration $t_{dur}$ occur on the sky at any one time, when $t_{dur}$ is on the order of minutes to hours.  For transients on the order of months to years, fewer than 160 ($t_{dur} / 1$ year) occur on the sky at any one time.   For transients on the order of years to millennia, fewer than 50 ($t_{dur}/1$ year)$^2$ exist.

\section{Discussion and Conclusions}
\label{conclusions}

An automated digital search was conducted for optical transients that would have been visible to the unaided eye.  The search lasted for three years, involved two CCD cameras, and thousands of fisheye frames.  Transients that brightened by more than three magnitudes to become visual magnitude 5.5 or brighter were detectable.  Only a single candidate optical transient was found.   It is therefore concluded that such transients on the order of minutes are rare, with $N_i < 0.004 \ (t_{dur} / 60$ seconds) transients on the sky at any one time.  Longer duration transients, however, are not so strongly excluded, with less than 160 transients of order one year appearing on the sky at any one time. 

Informal anecdotes about naked-eye transients seen by humans could be interpreted as indicating that bright transients could exist at a rate significantly in excess of that found here.  In that sense, the transient rate reported above is one of the first true limits on extremely bright transients.

Additionally, the above quoted limits do not appear to be in conflict with the rates of any known transient phenomena with the possible exceptions of comets.  As comets may also be a good test case, they are discussed specifically here.  The Astronomy Picture of the Day website \citep{Nem95} has reported that 12 comets that have become visible to the unaided eye since 1995.  These comets are Hyakutake in 1996, Hale-Bopp in 1997, LINEAR A2 in 2001, LINEAR WM1 in 2002, Ikeya-Zhang in 2002, LINEAR T7 in 2004, Bradfield in 2004, NEAT Q4 in 2004, Machholz in 2005, Pojmanski in 2006, McNaught in 2007, and Holmes in 2007.  

Only the comets listed for 2004 through 2006 occurred during our transience campaign.  Of these comets, only Comet Machholz and Holmes were detected on CONCAM frames. Comet Holmes was detected by eye after the transience campaign had ended \citep{ElH09}.  Comet Machholz, however, was detected repeatedly by the transience software, but not included because its origin was known.  

Both Comet LINEAR T7 and Comet Bradfield were either too dim or too low in the sky to trigger the automated search.   As some comets brighten only as they near the Sun, the later category is not a surprise.   Comet NEAT Q4 was too far south to trigger the northern Canary Island CONCAM when it was bright enough to have done so.  Comet Pojmanski was just too dim for CONCAMs to detect.

It is interesting to wonder if comets were not explicitly considered, would their transient rate have been in excess of that excluded by the CONCAM transience campaign?  The APOD comet statistics roughly translate into about one comet ``transient event" potentially detectable per year somewhere in the sky to our search paradigm.  Estimating that the average visibility time for a comet is about one week results in the total number of comet transients on the sky at any one time as $N_i \sim 0.019$.  Given the angular and temporal efficiency factors listed in Table 2 of $f_a \sim 0.33$ and $f_t \sim 0.25$ for a transient of one week duration, the grand total number of transients expected in this campaign would be $N_g \sim f_a f_t N_i \sim 0.0016$.  This number is significantly less than -- and so consistent with -- the upper limit of 19.1 transients found in the above analysis, listed in Table 2, and shown in Fig. 2.  

Besides OT060420 \citep{Sha06b}, no other unexplained transients satisfied the requirements of our automated transient detection algorithm.  On occasion, the locations of GRB afterglows were double checked by hand, but none were found or expected.  Contrastingly, variable stars Mira, $\chi$ Cygnus, and $\beta$ Lyra were detected by the automated software, but excluded by the authors because their origin was known.  

In the future, it would be possible for CONCAM-like devices to search for objects fainter than ``naked-eye", but this campaign did not do so.  Possibly the simplest methods would involve longer exposures and co-adding exposures.  Such methods would become significantly more difficult at the background brightness of an empty pixel \citep{NemToS03}, which corresponds to about visual magnitude 8 for the CONCAMs used in this campaign.  Other methods geared toward discovering fainter transients might increase the aperture size of the system or decrease the pixel size and hence the background noise.

\clearpage

\begin{figure}[h]
\includegraphics[width=\textwidth]{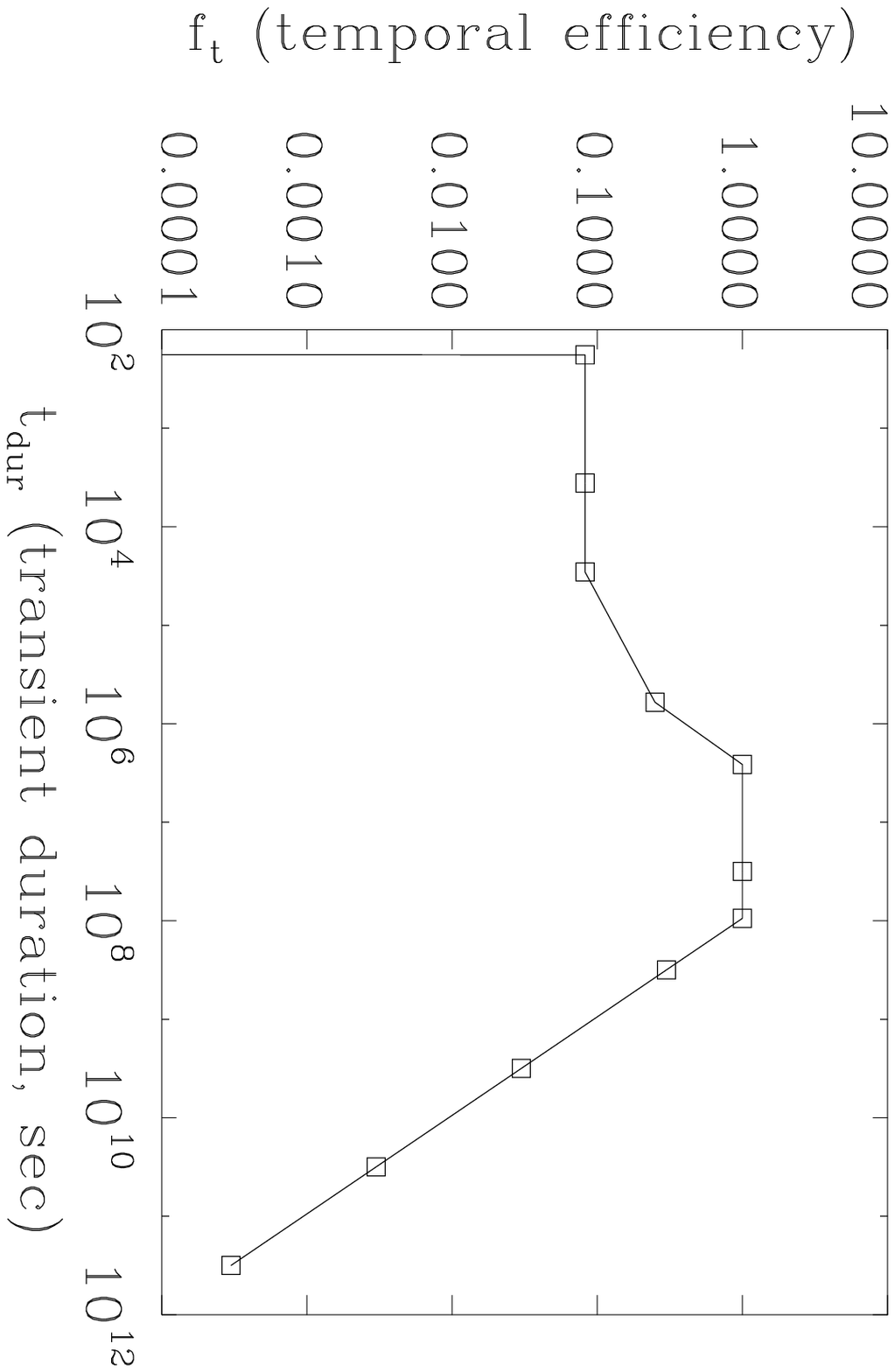}
\caption{The temporal efficiency of the three year transient search as a function of the duration of the transient. }
\label{Fig1}
\end{figure}

\clearpage
\begin{figure}[v]
\includegraphics[width=\textwidth]{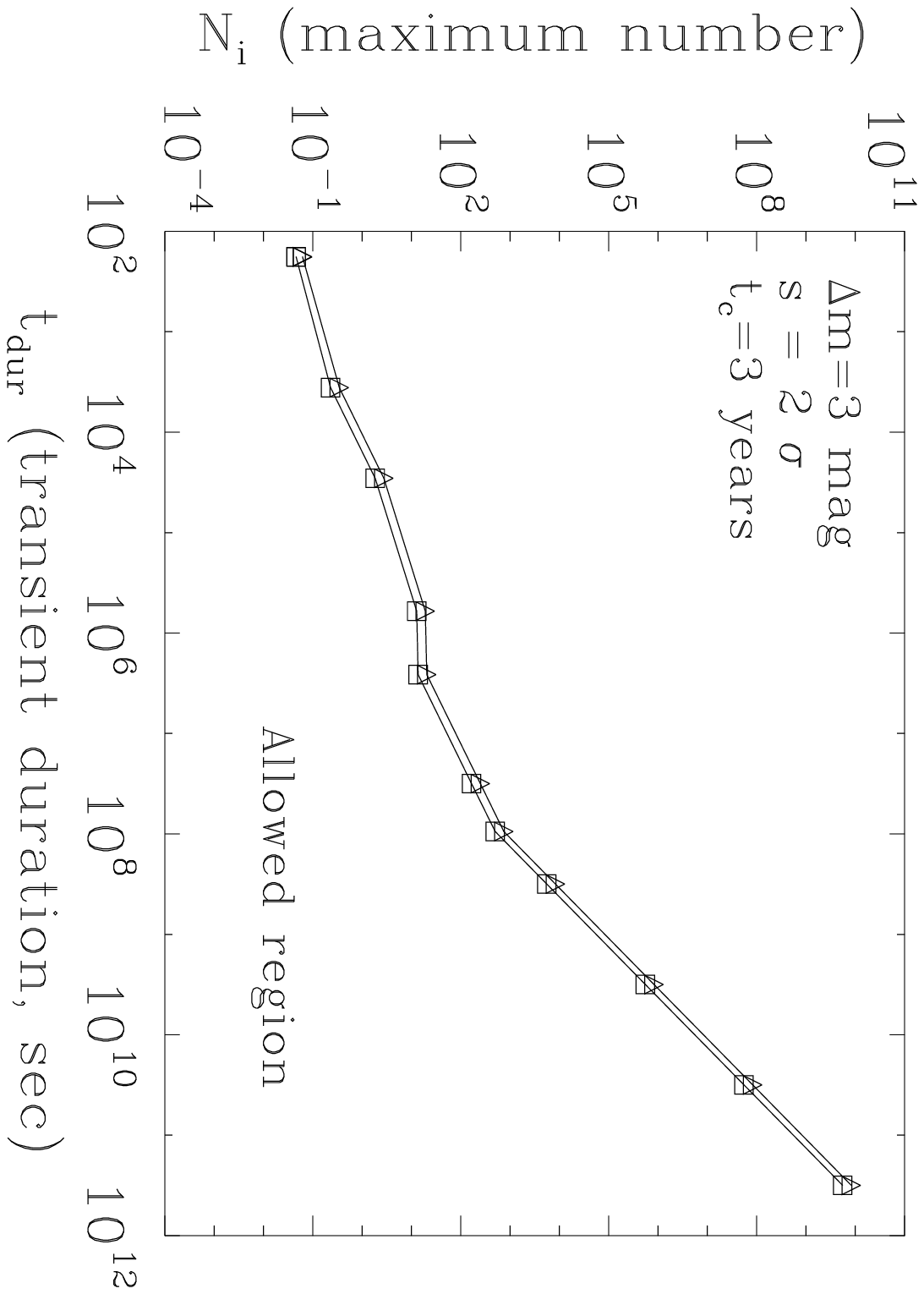}
\caption{The maximum number of instantaneous transients that appear anywhere on the sky at any given time. }
\label{Fig2}
\end{figure}

\end{document}